\documentclass[epj]{svjour}
\usepackage{latexsym}
\usepackage{graphicx}

\begin{document}
\title{Isospin and symmetry energy effects on nuclear fragment
production in liquid-gas type phase transition region}

\author{N. Buyukcizmeci\inst{1}, R. Ogul\inst{1,2} \and A.S. Botvina\inst{2,3}
}                     % Do not remove

\institute{Department of Physics, University of Sel\c{c}uk, 42075
Konya, Turkey.\and Gesellschaft fur Schwerionenforschung, D-64291
Darmstadt, Germany. \and Institute for Nuclear Research, Russian
Academy of Science, RU-117312 Moscow, Russia.}
\date{Received: date / Revised version: date}

\abstract{We have demonstrated that the isospin of nuclei
influences the fragment production during the nuclear liquid-gas
phase transition. Calculations for $Au^{197}$, $Sn^{124}$,
$La^{124}$ and $Kr^{78}$ at various excitation energies were
carried out on the basis of the statistical multifragmentation
model (SMM). We analyzed the behavior of the critical exponent
$\tau$ with the excitation energy and its dependence on the
critical temperature. Relative yields of fragments were classified
with respect to the mass number of the fragments in the transition
region. In this way, we have demonstrated that nuclear
multifragmentation exhibits a 'bimodality' behavior. We have also
shown that the symmetry energy has a small influence on fragment
mass distribution, however, its effect is more pronounced in the
isotope distributions of produced fragments.
  % end of PACS codes
} %end of abstract
\authorrunning{N. Buyukcizmeci, R. Ogul \and A.S. Botvina}
\titlerunning{Isospin and symmetry energy effects on nuclear fragment
production...}
\maketitle
\section{Introduction}

One of the most interesting phenomena in heavy ion reactions is
multifragmentation of nuclei, which is a very promising process to
determine the properties of nuclear matter at subnuclear densities
and high excitation energies. Experimental and theoretical
analysis of this phenomenon is also very important for our
understanding of astrophysical events, such as supernova
explosions and formation of neutron stars \cite{Bethe,ASBotvina}.
The ground state properties of nuclear matter can usually be
determined as a theoretical extrapolation, based on nuclear models
designed to describe the structure of real nuclei. When nuclei are
excited, these ground state properties show small changes without
structural effects. At very high excitation energies ($E^*>2-3$
MeV/nucleon) the nuclei can slowly expand by remaining in a state
of the thermodynamic equilibrium. During this expansion, they
enter the region of subsaturation densities, where they become
unstable to density fluctuations, and break up into the fragments
(droplet formation). In this case, it is believed that a
liquid-gas type phase transition is manifested. Even though there
exist some theoretical and experimental evidences of this phase
transition, so far some uncertainties remain in its nature
\cite{Goodman,Pethick,Ogul,Larionov,Agostino,Bauer,Hauger,Schmelzer,Mahi}.
Recently, the problem of isospin effects in multifragmentation
gains a lot of interest in connection with the astrophysical
applications. The main goal of this paper is to analyze these
effects in the framework of a realistic model.

\section{Statistical multifragmentation model}

A model used for extracting the properties of nuclear matter,
should be capable of reproducing the observed fragmentation
properties. Statistical multifragmentation model (SMM)
\cite{Botvina,Bondorf} has provided a good reproduction of
experimental data, as shown by many experimental analyses
\cite{Agostino,Hauger,Bondorf,Agosti96,Botvina1,Srivastava1,Karnaukhov}.
Within the SMM we consider the whole ensemble of breakup channels
consisting of hot fragments and nucleons in the freeze-out volume.
In finite nuclear systems the SMM takes into account the
conservation laws of energy $E^*$, momentum, angular momentum,
mass number $A_0$ and charge number $Z_0$. An advantage of SMM is
that it considers all break-up channels including the compound
nucleus and one can study the competition among them. In this
respect, there is a natural connection between multifragmentation
and traditional decay channels of nuclei as the evaporation and
fission at low excitation energies. On the other hand the SMM can
also address the thermodynamical limit, i.e. very large systems,
that makes it suitable for astrophysical applications
\cite{ASBotvina}. The probability of any breakup channel is
assumed to be proportional to its statistical weight, and the
channels are generated by the Monte Carlo method. The physical
quantities such as average mass and charge fragment yields,
temperature and its variance are calculated by running the
summations over all members of the ensemble.

In this work we use the SMM version described fully in
\cite{Bondorf}. However, for this paper, it is important to
explain how the hot primary fragments are treated. Light fragments
with mass number $A\le 4$ and charge number $Z\le 2$ are
considered as stable particles (``nuclear gas'') with masses and
spins taken from the nuclear tables. Only translational degrees of
freedom of these particles contribute to the entropy of the
system. Fragments with $A > 4$ are treated as heated nuclear
liquid  drops, and their individual free energies $F_{AZ}$ are
parameterized as a sum of the bulk, surface, Coulomb and symmetry
energy contributions
\begin{equation}
F_{AZ}=F^{B}_{AZ}+F^{S}_{AZ}+E^{C}_{AZ}+E^{sym}_{A,Z}.
\end{equation}
The bulk contribution is given by
$F^{B}_{AZ}=(-W_0-T^2/\epsilon_0)A$, where $T$ is the temperature,
the parameter $\epsilon_0$ is related to the level density, and
$W_0 = 16$~MeV is the binding energy of infinite nuclear matter.
Contribution of the surface energy is
$F^{S}_{AZ}=B_0A^{2/3}(\frac{T^2_c-T^2}{T^2_c+T^2})^{5/4}$, where
$B_0=18$~MeV is the surface coefficient, and $T_c=18$~MeV is the
critical temperature of the infinite nuclear matter. Coulomb
energy contribution is $E^{C}_{AZ}=cZ^2/A^{1/3}$, where $c$ is the
Coulomb parameter obtained in the Wigner-Seitz approximation,
$c=(3/5)(e^2/r_0)(1-(\rho/\rho_0)^{1/3})$, with the charge unit
$e$, $r_0$=1.17 fm, and $\rho_0$ is the normal nuclear matter
density $\sim 0.15fm^{-3}$. And finally, the symmetry term is
$E^{sym}_{A,Z}=\gamma (A-2Z)^2/A$, where $\gamma = 25$~MeV is the
symmetry energy parameter. All the parameters given above are
taken from the Bethe-Weizs\"acker formula and correspond to the
assumption of isolated fragments with normal density in the
freeze-out configuration. This assumption has been found to be
quite successful in many applications. However, for a more
realistic treatment one should consider the expansion of the
primary fragments in addition to their excitations. The residual
interactions among them should be considered as well. These
effects can be taken into account in the fragment free energies by
changing the corresponding liquid-drop parameters.

Here, we have applied the SMM model to nuclear sources of
different mass and isospin which can easily be used in modern
experiments. The excitation energy range is of $E^* = 2-20$
MeV/nucleon, and the freeze-out volume, where the intermediate
mass fragments are located after expansion of the system, is
$V=3V_0$ ($V_0 = 4\pi A_0 r_0^3/3$ is the volume of a nucleus in
its ground state). In other words, the fragment formation is
described at a low density freeze-out ($\rho \approx \rho_0/3$),
where the nuclear liquid and gas phases coexist. The SMM phase
diagram has already been under intensive investigations (see e.g.
\cite{bugaev}) with the same parametrization of the surface
tension of fragments versus the critical temperature $T_{c}$ for
the nuclear liquid-gas phase transition in infinite matter.
Therefore, it is possible to extract $T_{c}$ from the
fragmentation data \cite{Ogul1}. The symmetry energy contributes
to the masses of fragments as well. Taking this contribution into
account we shall investigate the influence of the symmetry energy
coefficient $\gamma$ on fragment production in the
multifragmentation of finite nuclei. In this paper we concentrate
mainly on properties of hot fragments. However, their secondary
deexcitation is important for final description of the data, which
was included in SMM long ago, and its effects were already
discussed somewhere else (\cite{Bondorf,Ogul1,Botvina87}). Here we
demonstrate how the secondary deexcitation code can be modernized
so that the change in the symmetry energy of nuclei during the
evaporation is taken into account.

\section{Isospin dependence of fragment distributions}

\begin{figure}
\begin{center}
\includegraphics[width=8.5cm,height=8.5cm]{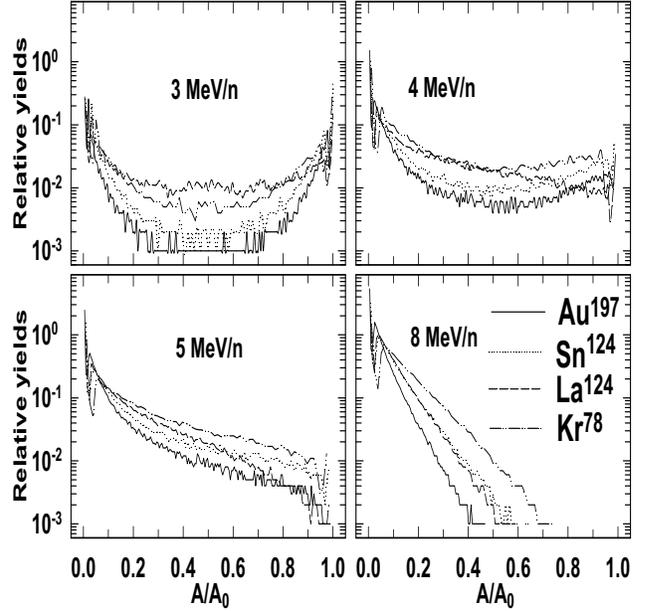}
\end{center}
\caption{Relative yield of hot primary fragments versus $A/A_{0}$
for $Au^{197}$, $Sn^{124}$, $La^{124}$ and $Kr^{78}$ at different
excitation energies $3$, $4$, $5$ and $8$ MeV/nucleon.}
\label{fig:1}
\end{figure}

For our calculations, we consider $Au^{197}$, $Sn^{124}$,
$La^{124}$ and $Kr^{78}$ nuclei to see how isospin affects the
multifragmentation phenomena with their neutron-to-proton ratios
$(N/Z)$, which read 1.49, 1.48, 1.18 and 1.17 respectively. The
two sources $Au^{197}$ and $Sn^{124}$ have nearly the same $N/Z$
ratios 1.49 and 1.48, respectively, but different mass numbers,
whereas $Sn^{124}$ and $La^{124}$ have the same mass numbers, but
very different $N/Z$ ratios 1.48 and 1.18, respectively. To see
the effect of the size of the nucleus in our study, we have also
included $Kr^{78}$, which has the lowest mass number among the
considered nuclei here, but its $N/Z$ ratio 1.17 is very close to
that of $La^{124}$. Relative yield of hot fragments produced after
the break-up of $Au^{197}$, $Sn^{124}$, $La^{124}$  and $Kr^{78}$
nuclei as a function of $A/A_0$ is given in Fig. 1 for $E^*=3$,
$4$, $5$ and $8$ MeV/nucleon. As was shown in previous analysis of
experimental data (see e.g.
\cite{Agostino,Hauger,Bondorf,Botvina1}) these results are fully
consistent with experimental observations. The effect of isospin
of different sources in liquid-gas phase transition region can be
clearly seen for different excitation energies. The mass
distribution of fragments produced in the disintegration of
various nuclei evolves with the excitation energy. As can be seen
in Fig. 1, for $Au^{197}$ and $Sn^{124}$ multifragmentation onset
is about $5$ MeV/nucleon and for $La^{124}$ and $Kr^{78}$ it is
about $4$ MeV/nucleon at standard SMM parameters so that U-shape
mass distributions disappear at these energies. We can also
describe this evolution with the temperature (see the caloric
curves below). At low temperatures ($T \leq 5$ MeV), there is a
U-shape distribution corresponding to partitions with few small
fragments and one big residual fragment. At high temperatures ($T
\geq 6$ MeV), the big fragments disappear and an exponential-like
fall-off is observed. In the transition region ($T \simeq 5-6$
MeV), however, one observes a transition between these two
regimes, which is rather smooth because of finiteness of the
systems.

In Fig. 2 we demonstrate N/Z ratios of the hot fragments produced
in the freeze-out volume of the same systems. One can see that
neutron content of hot fragments is only slightly smaller than
that of the whole system since the most of the neutrons are still
contained in fragments. It is also seen that the neutron richness
of fragments is increasing with their mass number \cite{Botvina2}.
This is a general behavior for nuclear systems in equilibrium,
where binding energy is influenced by interplay of the Coulomb and
the symmetry energy. For some systems one can observe specific
isotopic effects such as the increasing of the neutron richness of
intermediate mass fragments (with Z=3--20) with the excitation
energy (compare the results for 3 and 8 MeV/nucleon). This is
because after removing heavy fragments by increasing excitation
energy, their neutrons are accumulated into the smaller fragments
\cite{Botvina2}.

\begin{figure}
\begin{center}
\includegraphics[width=8.5cm,height=8.5cm]{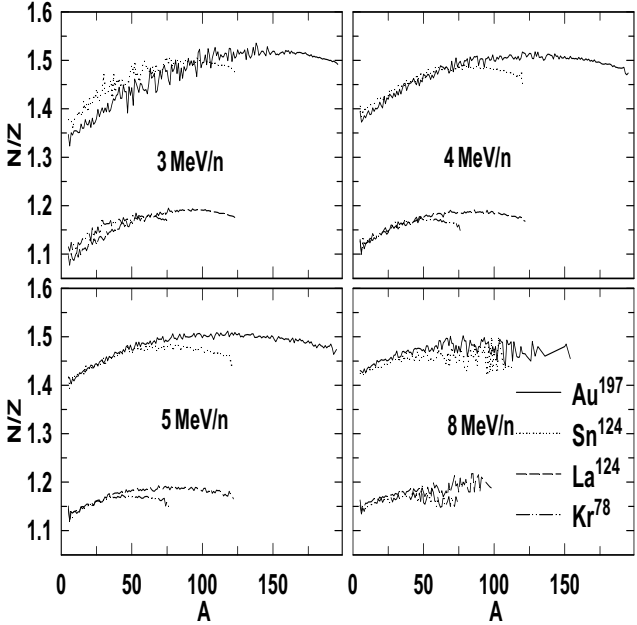}
\caption{N/Z ratio of hot primary fragments versus $A$ for
$Au^{197}$, $Sn^{124}$, $La^{124}$ and $Kr^{78}$ at different
excitation energies $3$, $4$, $5$ and $8$ MeV/nucleon.}
\end{center} \label{fig:2}
\end{figure}

Usually a power-law fitting is performed with, $Y(A) \propto
A^{-\tau}$ and $Y(Z) \propto Z^{-\tau_Z}$, where $Y(A)$ and $Y(Z)$
denote the multifragmentation mass and charge yield as a function
of fragment mass number $A$ and charge number $Z$, respectively.
The parameters $\tau$ (for mass distribution) and $\tau_Z$ (for
charge distribution) can be considered in thermodynamical limit as
critical exponents  \cite{Srivastava1,Reuter}. In calculations, we
consider the fragments in the range $6 \leq A \leq 40$  for mass
and $3 \leq Z \leq 18$ for charge number. The lighter fragments
are considered as a nuclear gas. The obtained values of $\tau$ and
$\tau_Z$ at $T_c = 18$ MeV as a function of $E^*$ are given in
Fig. 3 for cold fragmentation (with secondary deexcitation). There
is a minimum at about $E^*=5.3$, $5.2$, $4$ and $4.2$ MeV/nucleon
for $Au^{197}$, $Sn^{124}$, $La^{124}$ and $Kr^{78}$,
respectively. The results for $Au^{197}$ and $Sn^{124}$ are very
similar to each other and significantly different from those
obtained for $La^{124}$ and $Kr^{78}$. This means that
intermediate mass fragment (IMF) distributions approximately scale
with the size of the sources, and they depend on the
neutron-to-proton ratios of the sources. This is because the
symmetry energy still dominates over the Coulomb interaction
energy for these intermediate-size sources. One may also see from
these results that the lower N/Z ratio leads to smaller $\tau$ and
$\tau_Z$ parameters with the flatter fragment distribution. A
large N/Z ratio of the source favors the production of big
clusters since they have a large isospin. This is the reason for
domination of partitions consisting of small IMFs with a big
cluster in the transition region (U-shape distribution). It is
also seen from the bottom of Fig. 3 that the values of $\tau_Z$
are very close to the values of $\tau$ since the neutron-to-proton
ratio of produced IMFs exhibits a small change within their narrow
charge range (see also \cite{Botvina1,Botvina2}).

\begin{figure}
\begin{center}
\includegraphics[width=6.5cm,height=8cm]{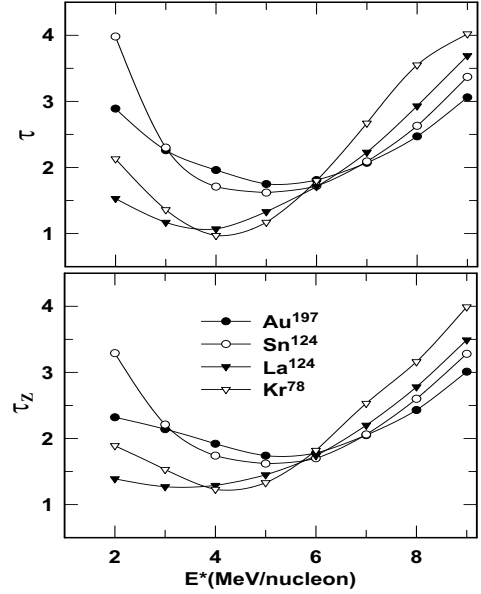}
\end{center}
\caption{The values of critical exponents $\tau$ (top panel) and
$\tau_Z$ (bottom panel), as a function of excitation energy $E^*$
for $Au^{197}$, $Sn^{124}$, $La^{124}$and $Kr^{78}$.}
\label{fig:3}
\end{figure}

\begin{figure}
\begin{center}
\includegraphics[width=7cm,height=8cm]{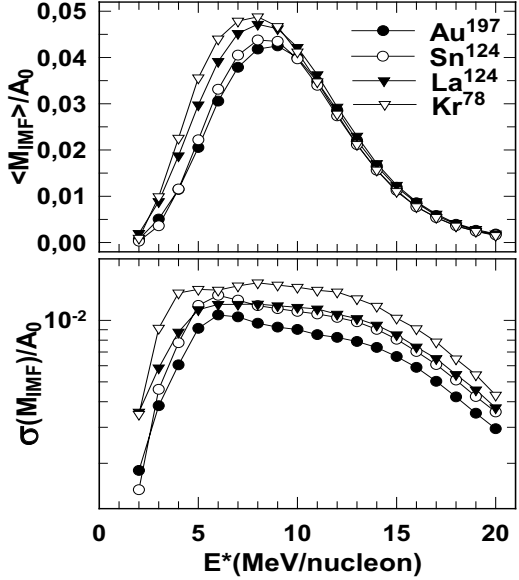}
\end{center}
\caption{The average IMF multiplicity (top panel) and its variance
(bottom panel) per nucleon versus $E^*$ for various nuclei.}
\label{fig:6}
\end{figure}

For completeness, we have calculated the average IMF multiplicity
and its variance for the excitation energy range of $2-20$
MeV/nucleon for all nuclei. To avoid the effect of the source size
and leave only the isospin effect, IMF multiplicity is divided by
$A_0$. In Fig. 4, we present the scaled average IMF multiplicity
and its variance versus $E^*$. In this figure, one may clearly see
an effect of the isospin and the source size on fragment
multiplicities.

\section{Largest fragments in fragment partitions}

In multifragmentation, the mass number of the largest fragment
$A_{max}$ may be used as an "order parameter" since it is directly
related to the number of fragments produced during disintegration.
We have analyzed its behavior with respect to the excitation
energies. In order to remove the effect of source size and leave
only an isospin effect, we have scaled it with the mass number of
the source $A_0$. We have plotted $A_{max}/A_{0}$ as a function of
excitation energy in Fig. 5 (top panel). The values of these
quantities decrease first rapidly with $E^{*}$, and then, beyond
some value ($E^{*}\sim$ 10 MeV/nucleon), decrease rather slowly.
This suggests that the vaporization process becomes dominant
around this point (see also an analysis in Ref.
\cite{Srivastava1}). Another striking result of this behavior is
that all curves almost coincide in the transition region as
displayed on the top panel in Fig. 5 (i.e. a universality of the
behavior of $A_{max}$). We have also calculated the average
$A_{max}$ variances in the multifragmentation stage for all nuclei
at the same excitation energy range. The average variance of
$A_{max}/A_{0}$ values are presented in the bottom panel of Fig.
5. We observe from this figure that the maximum variance values
for all sources are seen at an excitation energy of $4-5$
MeV/nucleon, which is exactly corresponding to the region of the
transition from compound-like channels to the full
multifragmentation. In the following we clarify why this kind of
behavior takes place.

\begin{figure}
\begin{center}
\includegraphics[width=7cm,height=8cm]{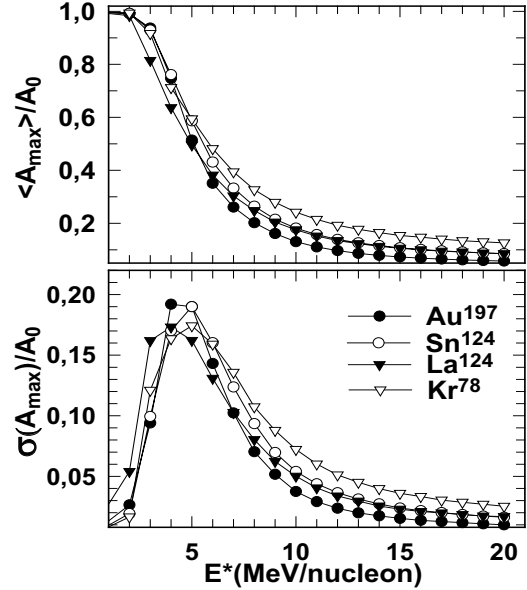}
\end{center}
\caption{$<A_{max}>$ and its variance $\sigma{(A_{max})}$ divided
by $A_0$ versus $E^*$ for $Au^{197}$, $Sn^{124}$, $La^{124}$ and
$Kr^{78}$.} \label{fig:4}
\end{figure}
\begin{figure}
\begin{center}
\includegraphics[width=7cm,height=14cm]{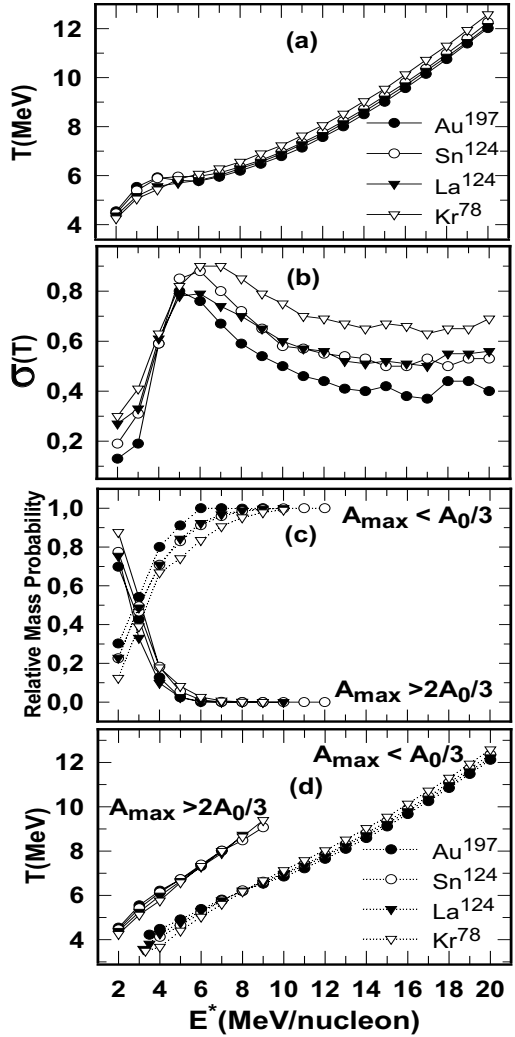}
\end{center}
\caption{(a) Temperature, (b) its variance value, (c)
probabilities of events with $A_{max}>2A_{0}/3$ and
$A_{max}<A_{0}/3$ and (d) average temperatures $T$  versus $E^{*}$
for $Au^{197}$, $Sn^{124}$, $La^{124}$ and $Kr^{78}$. The full and
dotted lines in Fig.6c and 6d represent the events with
$A_{max}>2A_{0}/3$ and $A_{max}<A_{0}/3$, respectively.}
\label{fig:5}
\end{figure}

\section{Temperature and bimodality}

Temperature of fragments is an important ingredient for any
description of nuclear multifragmentation. Besides statistical
approaches, a temperature can even be introduced in dynamical
(coalescence-like) processes of fragment production
\cite{Neubert03}. In SMM, the temperature of fragments $T_f$ in
separate partitions is defined from their energy balance according
to the canonical prescription:
\begin{equation}
m_{0}+E^{*}=1.5(n-1)T_f+\sum_i(m_i+E_{i}^{*}(T_f))+E_{coul},
\end{equation}
where $n$ is multiplicity of hot fragments and 'gas' particles,
$m_{0}$ and $m_{i}$ are ground masses of initial nucleus and the
fragments, $E_{i}^{*}$ are internal excitation energies of
fragments, and $E_{coul}$ is their Coulomb interaction energy.
Masses and internal excitations are found according to their free
energies (see eq.~(1) and Ref.~\cite{Bondorf}). The term with
$(n-1)$ takes into account the center of mass constraint, which is
important for finite systems \cite{Botvina03}. In this approach we
allow for temperature to fluctuate from partition to partition,
and we define the temperature of the system as the average one for
the generated partitions. The methods involving fluctuations of
temperature have been used in thermodynamics since long ago
\cite{Landau}. Our definition of temperature has another
advantage: It is fully consistent with the temperature commonly
used in nuclear physics for isolated nuclei, that makes a natural
connection with the physics of compound nucleus.

Figs. 6a and 6b show the results of the average temperature and
its variance with respect to the excitation energy $E^*$  for all
nuclei under consideration. While one may see a plateau in the
excitation energy range $3-7$ MeV/nucleon in the upper panel of
this figure, the variance of temperature exhibits a maximum value
at about $5-6$ MeV for standard SMM parameters, in agreement with
the maximum fluctuations of $A_{max}$. The plateau-like caloric
curve was reported long ago (e.g. \cite{Botvina,Bondorf}). It was
confirmed by experimental results \cite{Trautmann}, and by rather
sophisticated calculations of Fermionic Molecular Dynamics
\cite{Schnack} and Antisymmetrized Molecular Dynamics
\cite{Sugawa}. The reason for large fluctuations of the
temperature and $A_{max}$, can be clear from Figs. 6c and 6d. In
these figures we show relative mass probability and temperature
values for two groups of fragment partitions, which are defined as
$A_{max}>2A_{0}/3$ (associated with compound-like channels) and
$A_{max}<A_{0}/3$ (full multifragmentation).  As can be seen in
Fig. 6c, the probability of observing the first group events
$A_{max}> 2A_{0}/3$ decreases rapidly in the excitation energy
range $2-6$ MeV/nucleon, and this probability for the second group
$A_{max}<A_{0}/3$ increases within the same energy range. However,
temperatures of these groups of fragments are essentially
different, since the binding energy effect is different, and a
disintegration into a large number of fragments takes more energy.
The transition to the energy consuming group $A_{max}<A_{0}/3$
occurs because of the phase space domination, despite of a smaller
temperature of this group. We will call it 'bimodality' from now
on (see also discussion for other models in \cite{Chomaz}). It is
an intrinsic feature of the phase space population in the SMM,
however, it was only recently realized that it can be manifested
in different branches of the caloric curve and in momenta of
fragment distributions (see Ref. \cite{Agostino}). The bimodality
provides an explanation why the caloric curve may have a
plateau-like (or even a 'back-bending') behavior and large
fluctuations in the transition region.

\section{Influence of the critical temperature}

\begin{figure}
\begin{center}
\includegraphics[width=7cm,height=8cm]{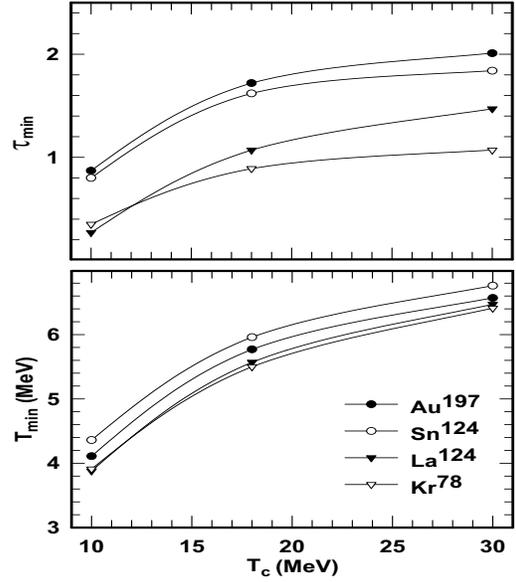}
\end{center}
\caption{The values of $\tau_{min}$ and $T_{min}$ as a function of
the critical temperature $T_c$ for various nuclei.}
\label{fig:7}
\end{figure}

We have analyzed how the critical exponent $\tau$ depends on
$T_c$, similar to Ref. \cite{Ogul1}, and on isospin of the
sources. During multifragmentation nuclei demonstrate critical
behavior at considerably lower temperatures ($T^{*}\approx 5-6$
MeV) than the critical temperature of nuclear matter $T_c$ (see
also \cite{Agostino,Hauger,Srivastava1}), because of Coulomb and
finite size effects. As was shown in many papers (see e.g.
\cite{Srivastava1} and references in) the critical behavior can
manifest as scaling of some observables around the critical point
$T^{*}$. In the present work we are interested in information
about the critical temperature $T_c$, which is high enough to be
extracted through the scaling in finite nuclei. However, this
information can be obtained by looking at other features of
fragment distributions, since the critical temperature influences
the surface tension of fragments. We have extracted the minimum
values of $\tau$ versus $T_c$ and the corresponding temperatures
of the systems, as shown in Fig. 7. The similar analysis of
experimental data was carried out by Karnaukhov et al.
\cite{Karnaukhov} for residues produced after intranuclear cascade
in Au target. One can see from Fig. 7 that both the $\tau_{min}$
and $T_{min}$ the temperature corresponding to this minimum,
increase with $T_c$. However, nuclei with lower isospin such as
$La^{124}$ and $Kr^{78}$ exhibit much lower $\tau_{min}$ values.
It is because of the proton-rich nuclei are less stable, and they
disappear rapidly with the excitation energy, this corresponds to
a less pronounced U-shape mass distribution. This effect may be
important for explanation of some data, since sources with
different isospin may manifest different $\tau_{min}$. The
decreasing $T_{min}$ for lower $T_c$ is explained by a very fast
decreasing of the surface tension with temperature in this case.
Nevertheless, all parameters increase with $T_c$, and tend to
saturate at $T_c \to \infty$, corresponding to the case of the
temperature-independent surface. In this case only the
translational and bulk entropies of fragments, but not the surface
entropy, influence the probability of fragment formation.

\section{Influence of the symmetry energy}

\begin{figure}
\begin{center}
\includegraphics[width=8cm,height=12cm]{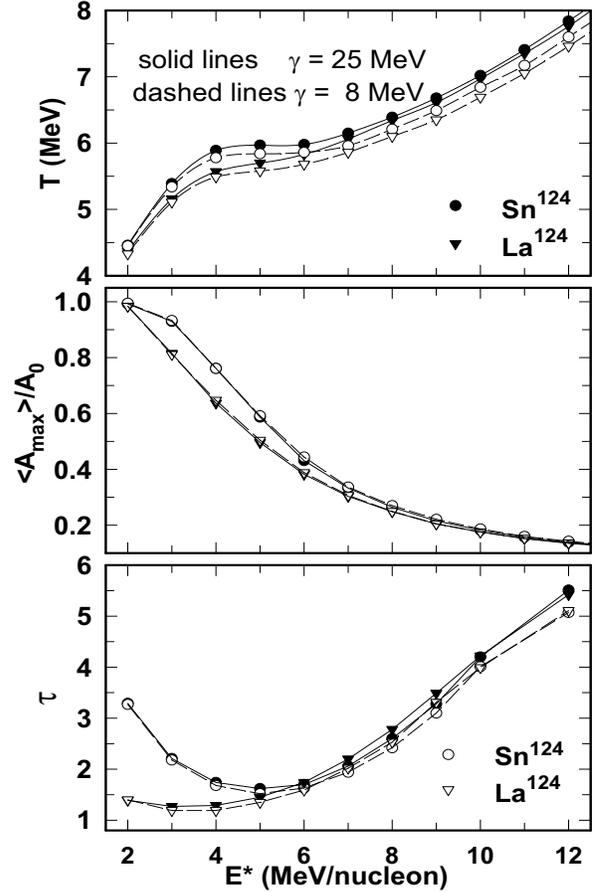}
\end{center}
\caption{Variation of the caloric curves (top panel),
$A_{max}/A_0$ (middle) and $\tau$ (bottom) of the nuclei
$Sn^{124}$ and $La^{124}$ with symmetry energy versus excitation
energy.}
\label{fig:8}
\end{figure}

Now let us turn to the analysis of the symmetry energy of
fragments, and its influence on fragment partitions. As we
discussed, the symmetry energy for fragments is defined in SMM as
$E^{sym}=\gamma (A-2Z)^2/A$, where $\gamma$ is a phenomenological
coefficient. As an initial approximation for the SMM calculations
we assume $\gamma$=25 MeV, corresponding to the mass formula of
cold nuclei. We have also performed the calculations for
$\gamma$=8 MeV for $Sn^{124}$ and $La^{124}$ sources. In Fig. 8 we
demonstrate the caloric curve, the mean maximum mass of fragments
in partitions, and $\tau$ parameters for different assumptions on
the symmetry energy. The results for $\gamma$=8 MeV are only
slightly different from those obtained for the standard SMM
assumption $\gamma$=25 MeV. From this figure, one can see a small
decrease of temperature caused by the involvement of the fragments
with unusual neutron and proton numbers into the dominating
partitions. Therefore, the average characteristics of produced hot
fragments are not very sensitive to the symmetry energy, and
special efforts should be taken in order to single out this
effect.

\begin{figure}
\begin{center}
\includegraphics[width=8cm,height=12cm]{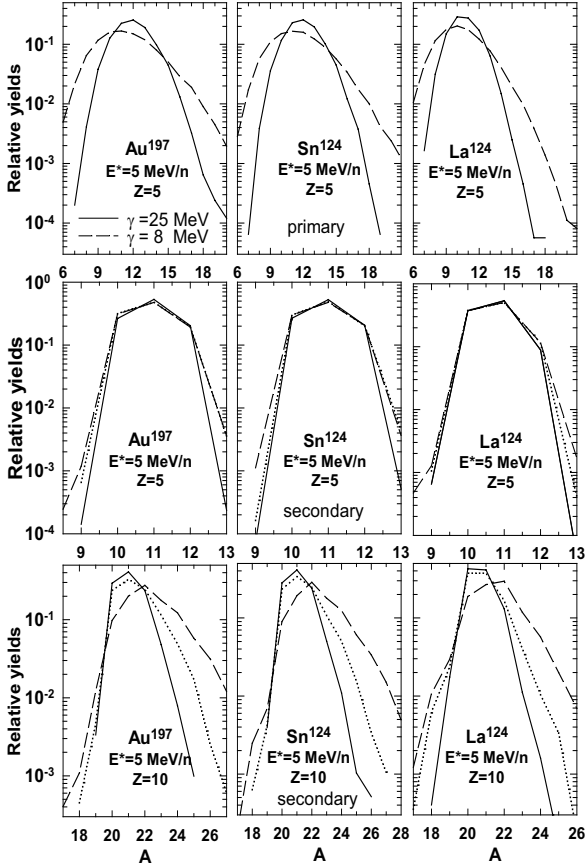}
\end{center}
\caption{ Isotopic distribution of primary and secondary fragments
for $Z=5$ and $Z=10$ versus mass number A at an excitation energy
of $5$ MeV/nucleon for $Au^{197}$, $Sn^{124}$ and $La^{124}$. At
$\gamma$=8 MeV the results after standard secondary decay are
shown with dotted lines, and dashed lines denote the results of
new evaporation depending on the symmetry energy(see the text).}
\label{fig:9}
\end{figure}

The main effect of the symmetry energy is manifested in charge and
mass variances of the hot fragments. In Fig. 9 we have shown the
relative mass distributions of primary hot (top panels) and final
cold fragments (after secondary deexcitation, middle and bottom
panels) with $Z=5$ and 10 at an excitation energy of $5$ MeV/n for
$\gamma =25$ and $8$ MeV. The distribution of primary hot
fragments becomes rather broad for $\gamma =8$ MeV and its maximum
lies on the neutron-rich side.

\section{Evaporation of hot fragments with isospin effects}

The primary fragments are supposed to decay by secondary
evaporation of light particles, or by the Fermi-break-up
\cite{Botvina87,Bondorf}. However, these codes use standard mass
formulae obtained from fitting masses of cold isolated nuclei. If
hot fragments in the freeze-out have smaller values of $\gamma$,
their masses in the beginning of the secondary deexcitation will
be different, and this effect should be taken into account in the
evaporation process. Sequential evaporation is considered only for
large nuclei ($A>$16) evaporating lightest particles
($n$,$p$,$d$,$t$,$^{3}$He,$\alpha$). We believe that we can
estimate the effect of the symmetry energy evolution during the
sequential evaporation by including the following prescription.
There are two different regimes in this process. First, in the
case when the internal excitation energy of this nucleus is large
enough ($E^{*}/A>1$MeV), we take for hot fragments the standard
liquid drop masses $m_{ld}$ adopted in the SMM, as follows
\begin{eqnarray}
\begin {array}{ll}
m_{AZ}=m_{ld}(\gamma)=&m_{n}N+m_{p}Z-AW_0+\gamma\frac{(A-2Z)^2}{A}\\
 &+B_0A^{2/3}+\frac{3e^2Z^2}{5r_0 A^{1/3}},
\end {array}
\end{eqnarray}
where $m_{n}$ and $m_{p}$ are masses of free neutron and proton.
In the second regime corresponding to the  lower excitation
energies, we adopt a smooth transition to standard experimental
masses with shell effects ($m_{st}$, taken from nuclear tables)
with the following linear dependence,
\begin{equation} \label{eq:mass}
m_{AZ}=m_{ld}(\gamma) \cdot x + m_{st} \cdot (1-x),
\end{equation}
where $x=\beta E^{*}/A$ ($\beta$= 1MeV$^{-1}$) and $x\leq$1. The
excitation energy $E^{*}$ is always determined from the energy
balance taking into account the mass $m_{AZ}$ at the given
excitation. This mass correction was included in a new evaporation
code developed on the basis of the old model \cite{Botvina87},
taking into account the conservation of energy, momentum, mass and
charge number. We have checked that the new evaporation at
$\gamma$=25 MeV leads to the results very close to those of the
standard evaporation.

Generally, the secondary deexcitation pushes the isotopes towards
the value of stability, however, the final distributions depend on
the initial distributions for hot fragments. One can see that the
results can also depend on whether the symmetry energy evolves
during the evaporation or not. In Fig. 9 we demonstrate the
results of two kind of calculations. First one is carried out
according to the standard code \cite{Botvina87} and the second one
is with the above described version taking into account the
symmetry energy (mass) evolution during evaporation. The standard
deexcitation leads to narrowing the distributions and to
concentration of isotopes closer to the $\beta-$stability line by
making the final distributions of the case  of $\gamma$=8 MeV
similar to the case of $\gamma$=25 MeV. However, a sensitivity to
the initial values of $\gamma$ remains, as one can see from
distributions of cold fragments with Z=5 and 10. This difference
in distributions is much more pronounced if we use the new
evaporation. The final isotope distributions are considerably
wider, and they are shifted to the neutron rich side, i.e. the
produced fragments are neutron rich. This effect has a simple
explanation: By using the experimental masses at all steps of
evaporation we suppress emission of charged particles by both the
binding energy and the Coulomb barrier. Whereas, in the case of
small $\gamma$ in the beginning of evaporation, the binding energy
essentially favors an emission of charged particles. When the
nucleus is cooled sufficiently down to restore the normal symmetry
energy, the remaining excitation is rather low ($E^{*}/A <$1 MeV)
for the nucleus to evaporate many neutrons.

The difference between the two kind of evaporation calculations
gives us a measure of uncertainty expected presently in the model.
This uncertainty can be diminished by comparison with experiments.
It is encouraging, however, that all final isotope distributions
have some dependence on the $\gamma$ parameter. This sensitivity
gives us a chance to estimate the symmetry energy by looking at
isotope characteristics. As was shown earlier the actual $\gamma$
parameter of the symmetry energy can be determined via an
isoscaling analysis \cite{Lozhkin}. An analysis of N/Z ratio of
produced fragments can also be used for this purpose. Recently one
may see some experimental evidences for essential decreasing of
the symmetry energy of fragments with temperature in
multifragmentation \cite{LeFevre,Shetty}. The consequences of this
decreasing are very important for astrophysical processes
\cite{ASBotvina}.

\section{Conclusions}

In summary, we have shown the effects of isospin, critical
temperature and symmetry energy on the fragment distribution in
multifragmentation. For this purpose we have analyzed different
nuclei with various neutron-to-proton ratio on the basis of SMM.
The critical temperature of nuclear matter influences the fragment
production in multifragmentation of nuclei through the surface
energy, while the symmetry energy determines directly the neutron
richness of the produced fragments. Effects of the critical
temperature can be observed in the power law fitting
parametrization of the fragment yields by finding $\tau$
parameter. We have also shown the effect of the isospin and
neutron excess of sources on the fragment distributions and on the
$\tau$ parametrization. By selecting partitions according to the
maximum fragment charge we have demonstrated a bimodality as an
essential feature of the phase transition in finite nuclear
systems. This feature may allow for identification of this phase
transition with the first order one. It has been found out that
the symmetry energy of the hot fragments produced in the
statistical freeze-out is very important for isotope
distributions, but its influence is not very large on the mean
fragment mass distributions after multifragmentation. We have
shown that the symmetry energy effect on isotope distributions can
survive after secondary deexcitation. An extraction of this
symmetry energy from the data is important for nuclear
astrophysical studies.

%\begin{acknowledgments}
N.B. and R.O. thank Selcuk University-Scientific Research Projects
(BAP) for a partial financial support under grant-SU-2003/033.The
authors gratefully acknowledge the enlightening discussions with
W. Trautmann, J. Lukasik, I.N. Mishustin, V.A. Karnaukhov and U.
Atav. We also thank GSI for hospitality, where part of this work
was carried out. R. Ogul gratefully acknowledges the financial
support by DAAD as well.
%\end{acknowledgments}


\begin{thebibliography}{27}
\bibitem{Bethe} H.A. Bethe, Rev. Mod. Phys. {\bf 62}, 801 (1990).
\bibitem{ASBotvina} A.S. Botvina and I.N. Mishustin,  Phys. Lett. B  {\bf 584}, 233 (2004).
\bibitem{Goodman}  A.L. Goodman, J.I. Kapusta and A.Z. Mekjian, Phys. Rev. C {\bf 30}, 851 (1984).
\bibitem{Pethick}  C.J. Pethick, D.G. Ravenhall, Nucl. Phys. {\bf A471}, 19c (1987).
\bibitem{Ogul}  R. Ogul, Int. J. Mod. Phys. E, {\bf 7(3)}, 419 (1998).
\bibitem{Larionov}  A.B. Larionov, I.N. Mishustin, Sov. J. Nucl. Phys. {\bf 57}, 636 (1994).
\bibitem{Agostino}  M. D'Agostino \emph{et al.}, Nucl. Phys. A {\bf 650}, 329 (1999).
\bibitem{Bauer} W. Bauer and A. Botvina, Phys. Rev.{\bf C52}, R1760 (1995).
\bibitem{Hauger}  J.A.Hauger \emph{et al.}, Phys. Rev.{\bf C62}, 024616 (2000).
\bibitem{Schmelzer}  J. Schmelzer \emph{et al.}, Phys. Rev. C {\bf 55}, 1917 (1997).
\bibitem{Mahi} M. Mahi \emph{et al.}, Phys. Rev. Lett. {\bf 60}, 1936 (1988).
\bibitem{Botvina}  A.S. Botvina, A.S. Iljinov, I.N. Mishustin, Sov. J. Nucl. Phys. {\bf 42}, 712 (1985).
\bibitem{Bondorf}  J.P. Bondorf, A.S. Botvina, A.S. Iljinov, I.N. Mishustin, and K. Sneppen, Phys. Rep. {\bf 257}, 133 (1995).
\bibitem{Agosti96}  M. D'Agostino \emph{et al.}, Phys. Lett {\bf B371}, 175 (1996).
\bibitem{Botvina1}  A.S. Botvina \emph{et al.}, Nucl. Phys. {\bf A584}, 737 (1995).
\bibitem{Srivastava1}  B.K. Srivastava \emph{et al.}, Phys. Rev. C {\bf 65}, 054617 (2002).
\bibitem{Karnaukhov}  V.A. Karnaukhov \emph{et al.}, Phys. Rev. C {\bf 67}, R011601 (2003).
\bibitem{bugaev} K.A. Bugaev {\it et al.}, Phys. Rev. {\bf C62}, 044320 (2000)
\bibitem{Ogul1}  R. Ogul and A.S. Botvina, Phys. Rev. C {\bf 66}, R051601 (2002).
\bibitem{Botvina87}  A.S. Botvina \emph{et al.}, Nucl. Phys. {\bf A475}, 663 (1987).
\bibitem{Botvina2} A.S. Botvina and I.N. Mishustin, Phys. Rev. C {\bf 63}, 061601 (2001).
\bibitem{Reuter}  P.T. Reuter, K.A.Bugaev, Phys. Lett. B {\bf 517}, 233 (2001).
\bibitem{Neubert03} W. Neubert and A.S. Botvina, Eur. Phys. J. A {\bf 17}, 559 (2003).
\bibitem{Botvina03} A.S. Botvina and I.N. Mishustin, Phys. Rev. Lett. {\bf 90}, 179201 (2003).
\bibitem{Landau} L.D. Landau and E.M. Lifshitz, Statistical Physics,
Part 1 (3rd ed. Pergamon, New York, 1980) p.340.
\bibitem{Trautmann}  W. Trautmann, Nucl. Phys. {\bf A685}, 233 (2001).
\bibitem{Schnack}  J. Schnack, H. Feldmeier, Phys. Lett. B {\bf 409}, 6 (1997).
\bibitem{Sugawa}  Y. Sugawa and H. Horiuchi, Phys. Rev. C {\bf 60}, 064607 (1999).
\bibitem{Chomaz}  Ph. Chomaz \emph{et al.}, Phys. Rev. E {\bf 64}, 046114 (2001).
\bibitem{Lozhkin} A.S. Botvina, O.V.Lozhkin, W. Trautmann, Phys. Rev. C {\bf 65}, 044610 (2002).
\bibitem{LeFevre} A.Le Fevre \emph{et al.}, Phys. Rev. Lett. {\bf 94}, 162701 (2005).
\bibitem{Shetty} D.V. Shetty \emph{et al.}, Phys. Rev. C {\bf 71}, 024602 (2005).

\end{thebibliography}
\end{document}